\documentclass[10pt]{article}
\pdfoutput=1
\usepackage[utf8]{inputenc}
\usepackage{graphicx}
\usepackage{geometry}
\usepackage{multirow}
\usepackage{caption}
\usepackage{times}
\usepackage{float}
\usepackage[usestackEOL]{stackengine}
\usepackage[table,xcdraw]{xcolor}

\title{Exploring Causes, Effects, and Solutions to Financial Illiteracy and Exclusion among Minority Demographic Groups}
\author{Abhinav Shanbhag}
\date{October 2022}

\begin{document}

\maketitle
\vspace*{-1cm}\section*{Abstract}
Americans across demographic groups tend to have low financial literacy, with low-income people and minorities at highest risk. This opens the door to the exploitation of unbanked low-income  families through high-interest alternative financial services. This paper studies the causes and effects of financial illiteracy and exclusion in the most at-risk demographic groups, and solutions proven to bring them into the financial mainstream. This paper finds that immigrants, ethnic minorities, and low-income families are most likely to be unbanked. Furthermore, the causes for being unbanked include the high fees of bank accounts, the inability of Americans to maintain bank accounts due to low financial assets or time, banking needs being met by alternative financial services, and being provided minimal help while transitioning from welfare to the workforce. The most effective solutions to financial illiteracy and exclusion include partnerships between nonprofits, banks, and businesses that use existing alternative financial service platforms to transition the unbanked into using products that meet their needs, educating the unbanked in the use of mobile banking, and providing profitable consumer credit products targeting unbanked families with features that support their needs in addition to targeted and properly implemented financial literacy programs.

\section{Introduction}
Over the years, the importance of financial literacy has grown tremendously throughout the world as financial respons- ibility is increasing. Today, many people have to pay off their student debt and mortgages, plan for their retirement, and manage investment accounts. Financial literacy ranges from budgeting and investing to taxes and insurance. However, while people are increasingly aware of the importance of financial literacy, financial literacy rates have remained relatively low, even in one of the most technologically advanced and educated countries in the world: the United States. A study conducted by The National Council on Economic Education was designed to evaluate adults’ and students’ understanding of basic economics as outlined in the Voluntary National Content Standards in Economics (Markow and Bagnaschi, 2005). 3,512 U.S. adults aged 18+ and 2,242 U.S. students in grades 9-12 completed the survey. Additionally, the data was weighted to represent the total U.S population of adults 18 and over and the total U.S population of 9th – 12th grade students. The results were disappointing as, although progress had been made over the past years, a majority of high school students did not understand basic concepts in economics that are integral to function financially in the real world. In fact, 28\% of adults and 60\% of high school students got an “F” on the economics quiz, while just 34\% of adults and 9\% of high school students got an “A” or “B”(Markow and Bagnaschi, 2005).

These results not only depict the lack of financial literacy and financial education in high schools, but also the inability of most high school students to make sound financial decisions in the real world after graduating high school. Without basic knowledge of financial concepts, people have little awareness of the importance of paying back debts and loans, planning for retirement, and managing mortgages. This lack of knowledge is extremely problematic as Americans have to learn financial concepts through failure, which leads many into chronic debt or financial instability. Table 1 details that economic understanding increases with age, which suggests that Americans tend to learn economic concepts through experience, rather than knowledge they gain during high school - an educational institution whose first priority is to prepare students for the real world (Markow and Bagnaschi, 2005).

\begin{table}[h]
\centering
\captionsetup[table]{skip=10pt}
\caption{Scores on the Financial Literacy Quiz by Age Group}            
\begin{tabular}{|ccccc|}
\hline
\multicolumn{5}{|c|}{Age}                                                                                                                                            \\ \hline
\multicolumn{1}{|l|}{}              & \multicolumn{1}{c|}{\textbf{18-34}} & \multicolumn{1}{c|}{\textbf{35-49}} & \multicolumn{1}{c|}{\textbf{50-64}} & \textbf{65+} \\ \hline
\multicolumn{1}{|c|}{A}             & \multicolumn{1}{c|}{12\%}           & \multicolumn{1}{c|}{18\%}           & \multicolumn{1}{c|}{18\%}           & 20\%         \\ \hline
\multicolumn{1}{|c|}{B}             & \multicolumn{1}{c|}{13\%}           & \multicolumn{1}{c|}{16\%}           & \multicolumn{1}{c|}{19\%}           & 22\%         \\ \hline
\multicolumn{1}{|c|}{C}             & \multicolumn{1}{c|}{25\%}           & \multicolumn{1}{c|}{23\%}           & \multicolumn{1}{c|}{24\%}           & 27\%         \\ \hline
\multicolumn{1}{|c|}{D}             & \multicolumn{1}{c|}{14\%}           & \multicolumn{1}{c|}{13\%}           & \multicolumn{1}{c|}{14\%}           & 13\%         \\ \hline
\multicolumn{1}{|c|}{F}             & \multicolumn{1}{c|}{35\%}           & \multicolumn{1}{c|}{30\%}           & \multicolumn{1}{c|}{25\%}           & 18\%         \\ \hline
\multicolumn{1}{|c|}{Average Score} & \multicolumn{1}{c|}{66}             & \multicolumn{1}{c|}{70}             & \multicolumn{1}{c|}{73}             & 76           \\ \hline
\end{tabular}
\end{table}

The percentages of Americans that received an “F” on the economics quiz, as reported by the study, are listed below: (Markow and Bagnaschi, 2005). \\*

\hspace*{0.19in}- 66\% of 9th - 10th graders got an “F” \\*
\hspace*{1cm}- 53\% of 11th - 12th graders got an “F” \\*
\hspace*{1cm}- 35\% of 18 - 34 year olds got an “F” \\*
\hspace*{1cm}- 30\% of 35 - 49 year olds got an “F” \\*
\hspace*{1cm}- 25\% of 50 - 64 year olds got an “F” \\*
\hspace*{1cm}- 18\% of 65+ year olds got an “F” \\*

Learning financial concepts through experience and failure can be detrimental to the lives of Americans. To learn about credit card debt, Americans have to experience the impact of racking up high debt through high-interest credit cards. To file taxes correctly, Americans have to experience facing penalties from the IRS for filing taxes incorrectly or failing to pay taxes entirely.  

The national financial literacy rate has been reported to have either declined, stayed stable, or grown slightly over the past 2 decades. The National Financial Capability Study in 2018 indicates that financial knowledge has been trending downwards. The study, in fact, found that respondents who answered four or more questions correctly declined from 44\% in 2015 to 40\% in 2018 (Lin et al. 2018). Some surveys, such as the Jump\$tart Coalition Survey of High School Seniors and College Students, done bi-annually by the Jump\$tart coalition for financial literacy, have reported that financial literacy rates even declined in the early 2000s (Mandell, 2008).  Multiple other studies have also documented the low financial literacy rate in the United States (Bernheim, 1995; Bernheim, 1998; Lusardi and Mitchell, 2007b; Lusardi and Tufano, 2008).

To improve the financial education high schoolers receive, national economic and finance organizations - the Council for Economic Education, for example - have pushed for laws and set national standards. The Council for Economic Education has set notable standards including the National Standards for Financial Literacy. Additionally, 45 states have included personal finance in their K-12 standards. The website of the National Conference of State Legislatures details various financial legislation in 2021 and 2022 as well (Morton and Lesley, 2021; Morton and Lesley, 2022). However, the effectiveness of financial education across the country has not been proven due to mixed results from various studies measuring the effectiveness of high school financial literacy and personal finance programs in high school. This may come across as surprising to many Americans, especially with the increasing awareness and legislation for financial literacy over the years. Low national financial literacy rates and the ambiguity in the effectiveness of high school financial education programs have prompted national organizations - committed to improving financial literacy and education - and the government to create new policies in an effort to mitigate financial illiteracy.

\begin{table}[h]
\centering
\captionsetup[table]{skip=10pt}
\caption{Financial Literacy Scores by Banked Status}
\begin{tabular}{|l|c|c|c|}
\hline
                    & \multicolumn{1}{l|}{\textbf{Full Sample}} & \multicolumn{1}{l|}{\textbf{Unbanked}} & \multicolumn{1}{l|}{\textbf{Banked}} \\ \hline
Financial Literacy Score (out of 5) & 3.007                                     & 2.026                                  & 3.062                                \\ \hline
Savings Question Correct            & 0.782                                     & 0.619                                  & 0.791                                \\ \hline
Inflation Question Correct          & 0.649                                     & 0.408                                  & 0.662                                \\ \hline
Bond Question Correct               & 0.278                                     & 0.195                                  & 0.283                                \\ \hline
Mortgage Question Correct           & 0.761                                     & 0.498                                  & 0.775                                \\ \hline
Stock Diversity Question Correct   & 0.537
   & 0.306   
& 0.55                                 \\ \hline
\end{tabular}
\end{table}

As this research paper focuses on both financial illiteracy and financial exclusion, it is important to show the correlation between both topics. The data set in Table 2 is used from the 2009 Financial Capability in the United States survey created by the Financial Industry Regulatory Authority (Lusardi, 2011). A State-By-State survey was used because of the large number of observations (28,146 American adults with a minimum of 500 adults from each state). The questionnaire asked respondents 5 questions relating to finance to gauge relative financial literacy rates in different demographic groups; unbanked respondents consistently scored lower than banked respondents on all 5 questions (Lusardi, 2011). The average score on the test for banked participants was a little more than 1 point higher than unbanked participants (Lusardi, 2011). The fact that unbanked participants have consistently lower scores goes to show that financial literacy does, in fact, reduce the likelihood of being unbanked. 

I am currently conducting financial literacy workshops with organizations in the Bay Area including but not limited to YMCA, Peninsula Bridge, and DreamCatchers, in addition to partnering with the parks and recreation departments of Bay Area cities such as Los Altos and Mountain View. This small-scale initiative is aimed at teaching integral personal finance concepts - budgeting, taxes, and credit to name a few - to middle school and high school students. Hands-on activities and online simulations have proved to have the largest impact on students’ education in personal finance. This research paper also serves the second purpose of investigating these observations further.  These data and corresponding studies fueled my motivation to further research the specific demographics at risk for being unbanked in addition to the causes, effects, and solutions to financial exclusion and illiteracy. This paper will be split into 4 categories, with each category diving deep into each of these topics.

\section{Demographics of Unbanked and Financially Illiterate Americans}

\begin{figure}[hbt!]
\centering
\captionsetup[figure]{skip=10pt}
\includegraphics[scale=0.4]{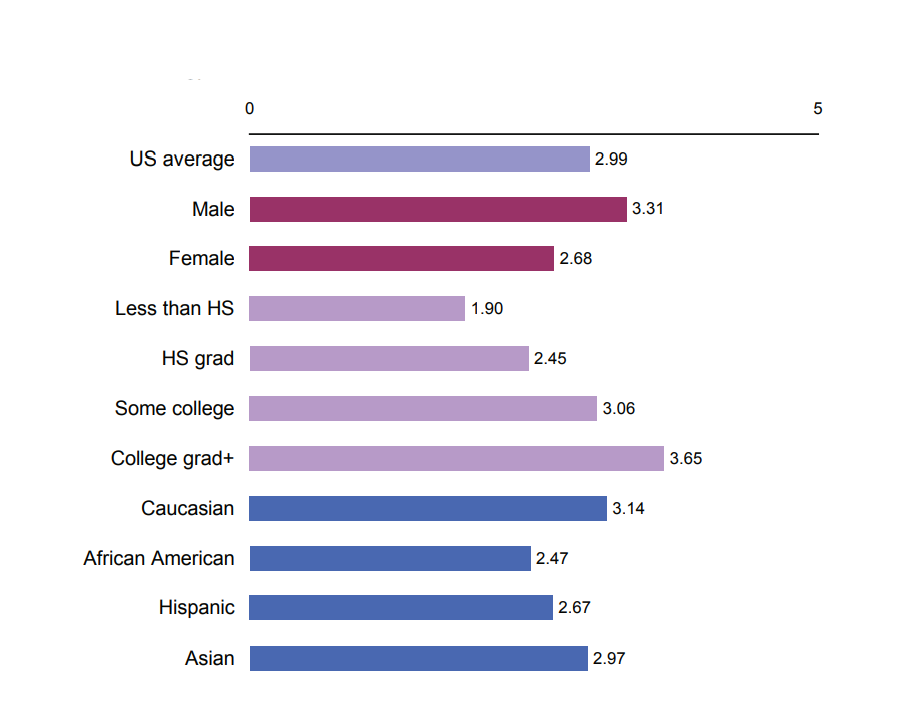}
\caption{Financial Literacy Index by Gender, Education, and Ethnicity}
\label{fig:3}
\end{figure}

\noindent The data from Figure 1 was collected from the National Financial Capability Study in 2009 (Lusardi, 2011). The data are average scores of certain demographic groups on a financial literacy quiz consisting of 5 questions. As shown in Figure 1, many demographic groups have low financial literacy. Average financial literacy scores are drastically lower among some demographic groups in the US, including women, those with lower education, and African-Americans and Hispanics. African-Americans scored an average of 0.7 points lower than Caucasians, while those with lower than a high school education scored an average of an astounding 1.75 points below college graduates (Lusardi, 2011). Although these scores are reported in single demographic characteristics, we can imagine Americans, who identify with multiple minority demographics (a female African-American with lower than a high school education for example), scoring much lower and thus being at higher risks of financial problems later down the road. 

These demographic groups are the most at risk groups for financial illiteracy, which can later be detrimental to their financial stability. These trends have been corroborated by many researchers (Lusardi and Mitchell, 2007, 2009; 2011a; Lusardi, Mitchell, and Curto, 2010; Lusardi and Tufano, 2009a,b). Therefore, solutions to relieve the burden of financial illiteracy need to remain a top priority if we wish to increase financial inclusion. 

Table 3 (data from a 2019 survey by the Federal Deposit Insurance Corporation (FDIC)) details which demographic groups are most at risk of being unbanked (Kutzbach et al. 2019). The definition we will use for unbanked will be “Adults who do not use or do not have access to any traditional financial services, including savings accounts, credit cards, or personal checks” (Downey, 2022). Demographic groups including low-income Americans (and their families), those with lower education, minority races, and younger Americans tend to be at a higher risk of being unbanked. The demographic groups shown to be most at-risk for being unbanked/financially illiterate, as per Figure 1 and Table 3, are the same. This implies that there is a correlation between certain demographic groups and being unbanked/financially illiterate. Americans with a family income less than \$15,000 are 50 times more likely to be unbanked than Americans with a family income of \$75,000 or larger (Kutzbach et al. 2019).  African-Americans are 6 times more likely than white Americans to be unbanked. This is significant because these differences highlight the stark contrast between minority groups and other demographic groups’ access to indispensable financial services and resources. It is also important to highlight the trends of unbanked rates from 2017 to 2019. Unbanked rates are undeniably getting better for most demographic groups, but a problem still remains: although unbanked rates have improved for minority races, they are still well above the unbanked rate of the white and high-income populations. In short, the change in the unbanked rates cannot be attributed to the impacts of targeted financial inclusion and literacy campaigns or laws but rather to the general growth in the standard of living in the United States. I am not stating that there has been no impact by non-profit organizations, the government, and financial institutions, but rather that the impact is not significant enough, especially since inclusion in the financial system is integral in this day and age.

\begin{table}[H]
\centering
\hspace{-0.5in}
\caption{Unbanked Rates by Household Characteristics and Year}
\begin{tabular}{|l|c|c|c|c|}
\hline
\rowcolor[HTML]{E7E3E3} 
\textbf{Characteristics}                                                             & \multicolumn{1}{l|}{\textbf{2015 (Percent)}} & \multicolumn{1}{l|}{\textbf{2017 (Percent)}} & \multicolumn{1}{l|}{\textbf{2019 (Percent)}} & \multicolumn{1}{l|}{\textbf{Difference (2019 - 2017)}} \\ \hline
All                                                                                  & 7.0                                          & 6.5                                          & 5.4                                          & -1.1                                                   \\ \hline
\textbf{Family Income}                                                               &                                              &                                              &                                              &                                                        \\ \hline
Less than \$15,000                                                                   & 25.6                                         & 25.7                                         & 23.3                                         & -2.5                                                   \\ \hline
\$15,000 to \$30,000                                                                   & 11.8                                         & 12.3                                         & 10.4                                         & -1.8                                                   \\ \hline
\$30,000 to \$50,000                                                                   & 5.0                                          & 5.1                                          & 4.6                                          & -0.5                                                   \\ \hline
\$50,000 to \$75,000                                                                   & 1.6                                          & 1.5                                          & 1.7                                          & 0.3                                                    \\ \hline
At least \$75,000                                                                    & 0.5                                          & 0.6                                          & 0.6                                          & 0.0                                                    \\ \hline
\textbf{Education}                                                                   &                                              &                                              &                                              &                                                        \\ \hline
No High School Diploma                                                               & 23.2                                         & 22.4                                         & 21.4                                         & -1.0                                                   \\ \hline
High School Diploma                                                                  & 9.7                                          & 9.4                                          & 8.1                                          & -1.4                                                   \\ \hline
Some College                                                                         & 5.5                                          & 5.1                                          & 4.3                                          & -0.9                                                   \\ \hline
College Degree                                                                       & 1.1                                          & 1.3                                          & 0.8                                          & -0.5                                                   \\ \hline
\textbf{Age Group}                                                                   &                                              &                                              &                                              &                                                        \\ \hline
15 to 24 Years                                                                       & 13.1                                         & 10.0                                         & 8.8                                          & -1.2                                                   \\ \hline
25 to 34 Years                                                                       & 10.6                                         & 8.5                                          & 6.9                                          & -1.6                                                   \\ \hline
35 to 44 Years                                                                       & 8.9                                          & 7.8                                          & 6.3                                          & -1.5                                                   \\ \hline
45 to 54 Years                                                                       & 6.7                                          & 6.9                                          & 5.1                                          & -1.8                                                   \\ \hline
55 to 64 Years                                                                       & 5.8                                          & 5.9                                          & 5.5                                          & -0.5                                                   \\ \hline
65 Years or More                                                                     & 3.1                                          & 3.9                                          & 3.3                                          & -0.6                                                   \\ \hline
\textbf{Race/Ethnicity}                                                              &                                              &                                              &                                              &                                                        \\ \hline
Black                                                                                & 18.5                                         & 16.8                                         & 13.8                                         & -2.9                                                   \\ \hline
Hispanic                                                                             & 16.3                                         & 14.4                                         & 12.2                                         & -2.2                                                   \\ \hline
Asian                                                                                & 3.9                                          & 2.6                                          & 1.7                                          & -1.0                                                   \\ \hline
\begin{tabular}[c]{@{}l@{}}American Indian or \\ Alaska Native\end{tabular}          & 15.3                                         & 18.0                                         & 16.3                                         & -1.7                                                   \\ \hline
\begin{tabular}[c]{@{}l@{}}Native Hawaiian or \\ Other Pacific Islander\end{tabular} & 10.3                                         & 2.8                                          & NA                                           & NA                                                     \\ \hline
White                                                                                & 3.1                                          & 3.0                                          & 2.5                                          & -0.6                                                   \\ \hline
Two or More Races                                                                    & 7.9                                          & 8.5                                          & 4.9                                          & -3.5                                                   \\ \hline
\textbf{Disability Status}                                                           &                                              &                                              &                                              &                                                        \\ \hline
Disabled, Aged 25 to 64                                                              & 17.6                                         & 18.1                                         & 16.2                                         & -1.9                                                   \\ \hline
Not Disabled, Aged 25 to 64                                                          & 6.5                                          & 5.7                                          & 4.5                                          & -1.1                                                   \\ \hline
\end{tabular}
\end{table}

Immigrants are another demographic group that is at a higher risk of being unbanked. Immigrants typically seek opportunities, employment, or education for their children in the United States. Access to financial services has numerous benefits for immigrant families. These include safety against theft or loss through depositing paychecks in transaction accounts in addition to bill paying, debit transactions, savings accounts for retirement, savings accounts for education/college, and electronic transfers. Electronic transfers are especially important for immigrants because they tend to send money to their families in their home country: this feat is risky and hard to accomplish without access to electronic transfers. Access to financial institutions enables immigrants to establish creditworthiness (important for convenience and debt consolidation), shield consumers from discriminatory or illegal lending practices through consumer protection laws, and provide help for managing personal/household finances (Rhine and Greene, 2006). The absence of these benefits leads immigrants to Alternative Financial Services (AFS) which can be disastrous. As shown in Table 4, immigrants are more likely to be unbanked than U.S born Americans (Rhine and Greene, 2006). Furthermore, Mexican immigrants are more likely than other immigrants (from Latin America, Asia, and Europe) to be unbanked. This is critical because Mexicans are the largest proportion of the immigrant population. In other words, a large number of immigrants are unbanked because they do not have the knowledge to make complex financial decisions regarding cashing paychecks, paying living expenses, and transferring money back home. With the continuous influx of immigrants into the United States, it is crucial to find solutions to transition immigrants into the financial mainstream in the United States or provide ways to increase their financial literacy. Otherwise, just like high school students that accumulate financial knowledge through failures, immigrants will also be forced to go into debt before learning about the financial system in the US (this is proven as the study details that immigrants that have stayed in the US longer are less likely to be unbanked than immigrants that have arrived recently).

\begin{table}[h]
\centering
\caption{Unbanked Rates by Immigration Status and Immigrant Group}
\begin{tabular}{|lc|}
\hline
\multicolumn{2}{|c|}{\textbf{Unbanked Rates}} \\ \hline
\multicolumn{1}{|l|}{U.S Born}       & 0.185  \\ \hline
\multicolumn{1}{|l|}{Immigrants}     & 0.323  \\ \hline
\multicolumn{1}{|l|}{Mexico}         & 0.533  \\ \hline
\multicolumn{1}{|l|}{Latin America}  & 0.372  \\ \hline
\multicolumn{1}{|l|}{Europe}         & 0.166  \\ \hline
\multicolumn{1}{|l|}{Asia}           & 0.198  \\ \hline
\end{tabular}
\end{table}

Finally, it is important to note that similar demographic groups tend to use alternative financial services (AFS) in place of traditional banking services. 91.3\% of White Americans with an income of at least \$75,000 used traditional banking services in comparison to the 45\% of white Americans with an income below \$15,000 that used said services (Kutzbach et al. 2019). Further, minority racial groups and white Americans with an income below \$15,000 have drastic differences in their percentage use of traditional banking services: 45\% for white Americans, 30.3\% for Hispanics, and 23.5\% for African-Americans. It is also important to note the differences in specific AFS use in regard to racial characteristics. In 2017, 31.4\% of African-Americans used non-bank money order services while only 9.1\% of white Americans did the same. In the same year, 11.5\% of African-Americans used check cashing services while only 4.6\% of white Americans did the same (Kutzbach et al. 2019). Finally, there is minimal change (or even a slight increase in some cases) from 2017 to 2019 regarding the use of AFS by at-risk demographic groups. For check cashing services, Americans with an income less than \$15,000 saw a 0.1\% increase while Americans without a high school diploma saw an increase of 3.1\% (Kutzbach et al. 2019). I have highlighted some data points to convey the fact that the initiative to bring unbanked Americans into the financial mainstream doesn’t have an impact that is significant enough. Data for other at-risk minority demographic groups, regarding their use of AFS, more or less follow the same pattern (a growing use of AFS or a slight decline).

\section{Causes of Being Unbanked or Financially Illiterate}

\begin{table}[ht]
\centering
\caption{Distribution of Reasons Cited by Respondents for not having a Checking Account}
\begin{tabular}{|l|l|l|l|l|}
\hline
\textbf{Reason}                                  & \textbf{2001} & \textbf{2004} & \textbf{2007} & \textbf{2010} \\ \hline
Do not write enough checks to make it worthwhile & 28.5          & 27.9          & 18.7          & 20.3          \\ \hline
Minimum balance is too high                      & 6.5           & 5.6           & 7.6           & 7.4           \\ \hline
Do not like dealing with banks                   & 22.6          & 22.6          & 25.2          & 27.8          \\ \hline
Service charges are too high                     & 10.2          & 11.6          & 12.3          & 10.6          \\ \hline
Cannot manage or balance a checking account      & 6.6           & 6.8           & 3.9           & 4.7           \\ \hline
Do not have enough money                         & 14.0          & 14.4          & 10.4          & 10.3          \\ \hline
Credit Problems                                  & 3.6           & *             & 6.6           & 4.2           \\ \hline
Do not need/want an account                      & 5.1           & 5.2           & 8.9           & 7.3           \\ \hline
Other                                            & 2.8           & 3.5           & 6.4           & 7.4           \\ \hline
Total                                            & 100           & 100           & 100           & 100           \\ \hline
\end{tabular} 
\begin{minipage}{13cm}
\hspace{0.3cm}
\small *Ten or fewer observations in any of the types of income
\end{minipage}
\end{table}

\noindent Table 5 is from a 2012 Federal Reserve Survey of Consumer Finances to show changes in family finances from 2007 to 2010 (Ackerman et al. 2012). The largest reason unbanked Americans do not want to use traditional banking services is that they have no reason to save, no time to save, or live paycheck to paycheck. In a study of the Unbanked and Underbanked Consumer in the Tenth Federal Reserve District in 2010 by the Federal Reserve Bank of Kansas City, unbanked and underbanked residents were asked about their experiences with being unbanked and why they choose to remain unbanked (Federal Reserve Bank, 2010). A common theme among unbanked residents was their limited, and often, unstable incomes. When Americans work paycheck to paycheck, they have virtually no time to save and plan for the future because it risks the survival of their family – they cannot afford to waste time because they wholly rely on their paychecks to survive. Additionally, unbanked and underbanked residents have limited use of bank accounts because many express a need for immediate access to their funds (Federal Reserve Bank, 2010). Their money will not stay in their bank account long and so they do not perceive a need to store money in a bank account, which they have to pay for, when they can store it at home for free. The greater risks of loss or theft from storing money at home were overshadowed by their need for immediate access to funds. An unbanked resident from Denver stated, “I wish I could have money to handle. I don’t have any. What I’m learning about myself is that I’m really trying to survive. Right now, I’m not working as much because of the economy, so I’m having a really hard time” (Federal Reserve Bank, 2010). With all of their money going towards living expenses, unbanked residents do not have any left over to cover the minimum balance requirements, which results in fees that most cannot afford. This lifestyle is indeed very risky because if an expensive emergency arises (car breaks down, unexpected debt collection, etc.), unbanked Americans have no way to pay without savings. 

\begin{table}[h]
\centering
\caption{Distribution of Assets by Age}
\begin{tabular}{|ccccc|}
\hline
\multicolumn{5}{|c|}{\textbf{Distribution of Assets (1997)}}                                                                                                                             \\ \hline
\multicolumn{1}{|c|}{}                                & \multicolumn{1}{c|}{Income Percentile} & \multicolumn{1}{c|}{Net Worth} & \multicolumn{1}{c|}{Financial Assets} & Housing Equity \\ \hline
\multicolumn{1}{|c|}{\multirow{5}{*}{Overall Sample}} & \multicolumn{1}{c|}{90}                & \multicolumn{1}{c|}{\$233,019} & \multicolumn{1}{c|}{\$82,638}         & \$129,415      \\ \cline{2-5} 
\multicolumn{1}{|c|}{}                                & \multicolumn{1}{c|}{75}                & \multicolumn{1}{c|}{\$110,846} & \multicolumn{1}{c|}{\$26,101}         & \$61,324       \\ \cline{2-5} 
\multicolumn{1}{|c|}{}                                & \multicolumn{1}{c|}{50}                & \multicolumn{1}{c|}{\$35,035}  & \multicolumn{1}{c|}{\$3,943}          & \$14,061       \\ \cline{2-5} 
\multicolumn{1}{|c|}{}                                & \multicolumn{1}{c|}{25}                & \multicolumn{1}{c|}{\$4,276}   & \multicolumn{1}{c|}{\$222}            & \$0            \\ \cline{2-5} 
\multicolumn{1}{|c|}{}                                & \multicolumn{1}{c|}{10}                & \multicolumn{1}{c|}{\$0}       & \multicolumn{1}{c|}{\$0}              & \$0            \\ \hline
\multicolumn{1}{|c|}{\multirow{5}{*}{Ages 25-34}}     & \multicolumn{1}{c|}{90}                & \multicolumn{1}{c|}{\$84,702}  & \multicolumn{1}{c|}{\$29,211}         & \$48,234       \\ \cline{2-5} 
\multicolumn{1}{|c|}{}                                & \multicolumn{1}{c|}{75}                & \multicolumn{1}{c|}{\$36,073}  & \multicolumn{1}{c|}{\$8,886}          & \$15,550       \\ \cline{2-5} 
\multicolumn{1}{|c|}{}                                & \multicolumn{1}{c|}{50}                & \multicolumn{1}{c|}{\$9,108}   & \multicolumn{1}{c|}{\$1,499}          & \$0            \\ \cline{2-5} 
\multicolumn{1}{|c|}{}                                & \multicolumn{1}{c|}{25}                & \multicolumn{1}{c|}{\$666}     & \multicolumn{1}{c|}{\$110}            & \$0            \\ \cline{2-5} 
\multicolumn{1}{|c|}{}                                & \multicolumn{1}{c|}{10}                & \multicolumn{1}{c|}{-\$1,633}  & \multicolumn{1}{c|}{\$0}              & \$0            \\ \hline
\end{tabular}
\end{table}

Looking closer at savings, Roth IRA and other savings accounts are either not accessible or provide minimal benefits for the unbanked and underbanked. Table 6 measures asset accumulation among low-income households (Carney and Gale, 2001). Americans in the lower 10th income percentile have zero net worth (some even have negative net worth), which means they have no way to contribute to savings accounts, traditional bank accounts, or investing accounts. What’s more surprising is that even those in the higher income percentiles (25\% and 50\%) still have relatively low financial assets. There are multiple potential reasons for this. One is that government policies discourage asset accumulation - the government imposes high implicit tax rates on asset accumulation. Another is that Americans have their financial assets tied up in housing equity - those in the 50th percentile have a significant amount of assets tied up in housing equity. A third reason is that Americans have refrained from saving - due to the two reasons mentioned above - which guarantees low financial assets in the future. Lastly, the cost of living have increased significantly. 

A study using data from the April 1993 Current Population Survey and its Survey of Employee Benefits supplement (CPS) detailed workers’ participation in 401(k) plans (Bassett et al. 1998). Among workers offered 401(k) plans, 35\% do not participate due to the fact that they do not receive any significant benefits from it. The data shows a pattern: workers with higher incomes are more likely to be offered 401(k) plans and make up a higher percentage of contributions to 401(k) plans. Additionally, participation rates in 401(k) plans for low-income workers are considerably higher when employers either offer them 401(k) plans or enroll them in one but give them the option to opt-out (Bassett et al. 1998). These results are very well-known – Richard Thaler won a Nobel prize for his theory to implement small psychological interventions called “nudges” to get more workers to invest in 401(k) plans. These results are relevant to this paper because it conveys that low-income workers cannot be expected to make the best financial decisions (either because they have no time or no knowledge of the cost/benefits of their decisions) and, thus, it is important to implement systems to “nudge” the unbanked and underbanked into making better financial decisions. Although we have made significant progress in the research of such support systems, more progress is needed in regards to employers implementing such systems because the unbanked and underbanked lack such systems in most areas of their lives.

Let’s look at other reasons as to why the unbanked do not benefit from the savings systems in place. Unbanked immigrants tend to transfer their savings to their families back home after paying their bills, leaving them with no savings. Thus, immigrants are less likely to have any savings to contribute, denying them the opportunity to reap the benefits many Americans enjoy.  Additionally, savings accounts provide tax benefits, which is really only an incentive for high-income Americans, who become richer. This follows the adage: the poor stay poor while the rich get richer. 

Another main reason the unbanked have cited for not having a bank account is that traditional bank accounts and savings accounts do not meet their needs.

\begin{table}[H]
\centering
\caption{Experience with Alternative Financial Services (AFS)}
\begin{tabular}{|lllll|}
\hline
\rowcolor[HTML]{E7E3E3} 
\multicolumn{1}{|l|}{\cellcolor[HTML]{E7E3E3}}                                                           & \multicolumn{1}{l|}{\cellcolor[HTML]{E7E3E3}\textbf{Full Sample}} & \multicolumn{1}{l|}{\cellcolor[HTML]{E7E3E3}\textbf{Fully Banked}} & \multicolumn{1}{l|}{\cellcolor[HTML]{E7E3E3}\textbf{Underbanked}} & \textbf{Unbanked} \\ \hline
\multicolumn{1}{|l|}{\textbf{Credit}}                                                                    & \multicolumn{1}{l|}{}                                             & \multicolumn{1}{l|}{}                                              & \multicolumn{1}{l|}{}                                             &                   \\ \hline
\multicolumn{1}{|l|}{Use payday loan ever}                                                               & \multicolumn{1}{l|}{11.2}                                         & \multicolumn{1}{l|}{6.0}                                           & \multicolumn{1}{l|}{42.6}                                         & 15.5              \\ \hline
\multicolumn{1}{|l|}{\begin{tabular}[c]{@{}l@{}}Use payday loan in \\ last 12 months\end{tabular}}       & \multicolumn{1}{l|}{29.9}                                         & \multicolumn{1}{l|}{*}                                             & \multicolumn{1}{l|}{64.2}                                         & 16.3              \\ \hline
\multicolumn{1}{|l|}{Use auto title loan}                                                                & \multicolumn{1}{l|}{3.6}                                          & \multicolumn{1}{l|}{*}                                             & \multicolumn{1}{l|}{29.5}                                         & *                 \\ \hline
\multicolumn{1}{|l|}{Use layaway}                                                                        & \multicolumn{1}{l|}{3.8}                                          & \multicolumn{1}{l|}{*}                                             & \multicolumn{1}{l|}{28.8}                                         & 5.4               \\ \hline
\multicolumn{1}{|l|}{\textbf{Payments}}                                                                  & \multicolumn{1}{l|}{}                                             & \multicolumn{1}{l|}{}                                              & \multicolumn{1}{l|}{}                                             &                   \\ \hline
\multicolumn{1}{|l|}{Use check casher}                                                                   & \multicolumn{1}{l|}{4.1}                                          & \multicolumn{1}{l|}{*}                                             & \multicolumn{1}{l|}{26.8}                                         & 10.1              \\ \hline
\multicolumn{1}{|l|}{Prepaid cards}                                                                      & \multicolumn{1}{l|}{}                                             & \multicolumn{1}{l|}{}                                              & \multicolumn{1}{l|}{}                                             &                   \\ \hline
\multicolumn{1}{|l|}{Gift card}                                                                          & \multicolumn{1}{l|}{48.0}                                         & \multicolumn{1}{l|}{51.5}                                          & \multicolumn{1}{l|}{48.8}                                         & 22.0              \\ \hline
\multicolumn{1}{|l|}{\begin{tabular}[c]{@{}l@{}}General-purpose \\     card\end{tabular}}                & \multicolumn{1}{l|}{14.5}                                         & \multicolumn{1}{l|}{13.2}                                          & \multicolumn{1}{l|}{17.9}                                         & 20.6              \\ \hline
\multicolumn{1}{|l|}{Payroll card}                                                                       & \multicolumn{1}{l|}{1.7}                                          & \multicolumn{1}{l|}{*}                                             & \multicolumn{1}{l|}{8.4}                                          & 6.6               \\ \hline
\multicolumn{1}{|l|}{Government card}                                                                    & \multicolumn{1}{l|}{4.8}                                          & \multicolumn{1}{l|}{3.2}                                           & \multicolumn{1}{l|}{6.0}                                          & 14.8              \\ \hline
\multicolumn{1}{|l|}{None}                                                                               & \multicolumn{1}{l|}{45.4}                                         & \multicolumn{1}{l|}{45.0}                                          & \multicolumn{1}{l|}{40.2}                                         & 54.6              \\ \hline
\multicolumn{1}{|l|}{\begin{tabular}[c]{@{}l@{}}Reloaded prepaid card \\ in last 12 months\end{tabular}} & \multicolumn{1}{l|}{59.7}                                         & \multicolumn{1}{l|}{33.3}                                          & \multicolumn{1}{l|}{53.7}                                         & 65.1              \\ \hline
\multicolumn{1}{|l|}{Most recent reload}                                                                 & \multicolumn{1}{l|}{}                                             & \multicolumn{1}{l|}{}                                              & \multicolumn{1}{l|}{}                                             &                   \\ \hline
\multicolumn{1}{|l|}{Past 7 days}                                                                        & \multicolumn{1}{l|}{21.2}                                         & \multicolumn{1}{l|}{24.2}                                          & \multicolumn{1}{l|}{*}                                            & *                 \\ \hline
\multicolumn{1}{|l|}{Past 30 days}                                                                       & \multicolumn{1}{l|}{41.1}                                         & \multicolumn{1}{l|}{35.4}                                          & \multicolumn{1}{l|}{53.3}                                         & 44.6              \\ \hline
\multicolumn{1}{|l|}{Past 90 days}                                                                       & \multicolumn{1}{l|}{20.0}                                         & \multicolumn{1}{l|}{18.3}                                          & \multicolumn{1}{l|}{28.8}                                         & *                 \\ \hline
\multicolumn{1}{|l|}{Past 12 months}                                                                     & \multicolumn{1}{l|}{17.1}                                         & \multicolumn{1}{l|}{21.1}                                          & \multicolumn{1}{l|}{*}                                            & *                 \\ \hline
\multicolumn{1}{|l|}{\begin{tabular}[c]{@{}l@{}}More than 12 \\     months ago\end{tabular}}             & \multicolumn{1}{l|}{*}                                            & \multicolumn{1}{l|}{*}                                             & \multicolumn{1}{l|}{*}                                            & *                 \\ \hline
\multicolumn{5}{|l|}{* Ten or fewer observations}                                                                                                                                                                                                                                                                                        \\ \hline
\end{tabular}
\end{table}

Table 5 shows that 28.5\% of unbanked people say that they do not write enough checks to make bank accounts worthwhile, 6.5\% say that the minimum balance is too high, and 22.6\% say that they do not like dealing with banks (Ackerman et al. 2012). Overall, it appears that unbanked Americans broadly do not think that bank accounts meet their needs. Alternative financial services (AFS), however, fill this need for unbanked Americans. Table 7 is from a survey by the Federal Reserve that explores the use of financial services by the underbanked and unbanked and the potential for mobile financial services adoption (Gross et al. 2012). According to this survey, two-fifths of underbanked households had used a payday loan; of these, two-thirds had used one within the past 12 months (Gross et al. 2012). Prepaid cards are becoming increasingly popular with underbanked and unbanked Americans - one-fifth of both underbanked and unbanked Americans use general-purpose prepaid cards, for example. Additionally, 53.7\% of underbanked Americans and 65.1\% of unbanked Americans have reloaded their pre-paid cards in the last year. We can conclude that AFS meet the needs of the unbanked. AFS are relatively easy to obtain, good in the short-term (with virtually no barriers to take out these loans, unbanked consumers can receive money during times of need), and have deceptive interest rates (consumers inaccurately believe that interest rates are low in the long-term). A 2012 survey by the Federal Reserve System on Consumers and Mobile Financial Services listed the main reasons why the unbanked use AFS instead of bank accounts (The Federal Reserve System, 2012).  The main reasons included thinking that they wouldn’t qualify for a bank loan or credit card, the relative ease of getting a payday loan than qualifying for a bank loan, and the fact that they can receive payday loans much faster. This is important because most unbanked Americans use AFS not because they believe AFS are better than traditional banking services but rather because they are denied access to banks - they do not have a choice.

This is a severe problem as AFS providers prey on unbanked Americans with deceptive interest rates and expensive extensions. They suffer because of debt or credit problems, either a reason for them to use AFS or a symptom of using them, and don’t come back out - they are in never-ending poverty and debt cycles. Kansas residents stated that they turned to retailers before banks due to simpler identification requirements, and more transparent fees, among others (Federal Reserve Bank, 2010). One resident said, “I cash my checks at Walmart because I can make one stop, and then if I want to send money to anybody, their [service] is right there. Everything is right there … It’s like just one stop” (Federal Reserve Bank, 2010). They noted that they were more confident in using these services - they were less likely to make mistakes - than banking services. 

Bank account fees are also an important cause of people being unbanked. In 1996, John Caskey interviewed 900 lower-income households on their use of financial services (Caskey, 1997b).  23.1\% of households said that they didn’t have deposit accounts because bank account fees were too high while 22.1\% said that banks require too much just to open an account (Caskey, 1997b). Unbanked Americans are dissuaded from having traditional bank accounts by overdraft fees and “hidden” fees  - banks make hefty profits from fees and so many believe that they are targeted by banks for profit.

To illustrate the evidence for such beliefs, let’s look at the high-to-low transaction reordering for consumers with checking accounts (Maggio et al. 2020).  Let’s say a customer has \$400 in her checking account. Her \$50 electric bill is deducted via automatic payment. During the day, she spends \$50 on groceries. At the end of the day, she has to pay \$5000 for rent. The customer’s bank charges \$35 in overdraft fees and under chronological transaction ordering (her cash from her bank account is deducted chronologically), she would pay \$35 once. However, under high-to-low transaction reordering, she pays in order of the highest expense (rent) to the lowest expenses (groceries/electric bill), which means she incurs 3 overdraft fees (\$105). The rent makes her balance negative, and she ends up having to pay more than she should in fees. Such bank policies target poorer consumers, ultimately disincentivizing them from opening bank accounts. Furthermore, as per Table 5, reasons like “Minimum Balance is too High” and “Service Charges are too High” haven’t seen much change over a period of 9 years (Ackerman et al. 2012). This is important to note because banks do not seem to want to help with the inclusion of the unbanked into the financial mainstream because it would risk a major revenue stream. 

It is not surprising that many unbanked Americans place the blame of not having a bank account on banks themselves. Table 5, in fact, states that 22.6\% of those surveyed did not like dealing with banks (Ackerman et al. 2012). In addition to “hidden” and overdraft fees, negative experiences can reduce Americans’ satisfaction with their banks. In the Kansas study, respondents felt at a disadvantage when trying to address issues related to bank-assessed fees - they did not feel “listened to'' or “believed” because of factors such as their lower income, manner of dressing, or language (Federal Reserve Bank, 2010). Banks communicated policies and fees poorly and did not take steps to increase convenience for certain customers. An underbanked woman in Kansas City stated, “Sometimes, the bank does not explain well the situation to us and then we make mistakes, and we have to pay for those mistakes. I feel that they should give us more time to be informed of the things that they are offering us” (Federal Reserve Bank, 2010).  For example, Hispanic immigrants considered the identification requirement process a barrier to opening a bank account. They stated that acceptable forms of identification were often unclear and that it could be different from bank to bank (Federal Reserve Bank, 2010). Imagine arriving in a new country - where the people, culture, and financial systems are different from one’s home country – and, instead of being offered much-needed support and aid, being looked down upon and denied any form of support. Previous studies that have gauged interest in having bank accounts among the unbanked found that around 70\% of those unbanked were not interested in having bank accounts; we can only assume that this is a result of the aforementioned negative experiences with banks (Kutzbach et al. 2019). Undocumented immigrants may also choose not to open bank accounts to keep their financial information private, either because they might fear that a bank record would reveal their presence to the Immigration and Naturalization Service or that their welfare eligibility could be threatened by a history of deposits from under-the-table earnings (Sherraden, 2005). With undocumented immigrants increasingly arriving in the U.S, it’s crucial for banks to have some sort of support system in place to help store their cash while they go through the procedures to obtain documents legally.

Finally, welfare families transitioning into the workforce are highly likely to be unbanked because of low financial knowledge and minimal help. A study by Edin and Lein on how welfare families manage to survive financially indicated that some employment opportunities and other avenues for financial growth were cut off by families’ lack of assets or chronic debt (Edin and Lein, 2016). Their debt or poor credit led these families to AFS providers, where financial charges exceeded the price of the goods sold. In short, when families previously on welfare left to become financially independent, they were very likely to be unbanked and forced to use AFS. Poverty was either a cause for or symptom of turning to AFS. When families depend on welfare, they have their finances taken care of which leads to them having low financial literacy and making bad financial decisions when they are financially independent. A study by Stegman and Faris, on the use of consumer credit by current and former TANF recipients in Charlotte, reports that around 50\% of TANF (temporary assistance for needy families) households are unbanked (Stegman and Faris, 2005).  Although leavers (families who left welfare in the 24 months prior to the survey and on the path to greater economic security) are better off, they are still relatively vulnerable as they now have to be financially independent without any financial knowledge, experience, or support (Stegman and Faris, 2005). In essence, it’s like being thrown into an ocean without learning how to swim. 27\% of these families are unbanked (Stegman and Faris, 2005).  The already difficult transition from cash assistance to the workforce can be especially detrimental when posed with additional challenges: saving money is difficult, they have less time to manage finances, paying monthly bills is time-consuming, and investing for retirement is all but impossible. Further, the study reports that only 38.5\% of families on welfare and 23.3\% of leavers pay off their balance every month. Additionally, 55.4\% of leavers use payday loans one or two times a year (Stegman and Faris, 2005). Without support systems in place to help families on welfare transition to the workforce, families have to go back on welfare, face chronic debt, or worse - they are forced into dire poverty with no way out.

\section{Effects of Being Unbanked or Financially Illiterate}

\begin{figure}[hbt!]
\centering
\hspace*{-1cm}\includegraphics[scale=0.6]{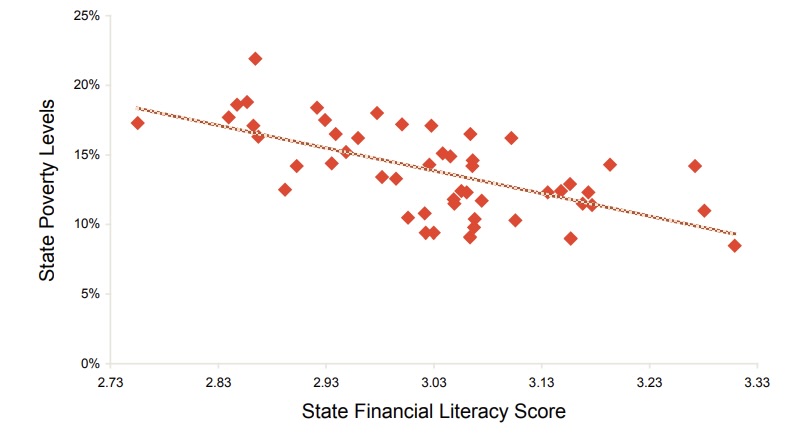}
\caption{Financial Literacy and Poverty Levels}
\label{fig:3}
\end{figure}

\noindent Figure 2 shows that financial literacy and poverty are highly correlated across states (Bumcrot, 2011). This graph does not try to prove causation in any way but solely aims to show the correlation between being financially illiterate and being in poverty. The poverty cycle entails that if Americans start off in poverty, they are less likely to receive equal access to financial literacy resources. Similarly, if Americans start off financially illiterate, their poor financial decisions, which are products of their financial illiteracy, will lead to chronic debt, ultimately pushing them into poverty. It’s important to understand the implications of this graph because solutions aimed at increasing financial inclusion must take both of these factors into account - one is rarely without the other. 

\begin{table}[H]
\centering
\captionof{table}[foo]{The Alternative Financial Service Industry}
\begin{tabular}{|l|l|l|l|l|}
\hline
\rowcolor[HTML]{E7E3E3} 
\textbf{Service}                                                       & \textbf{\begin{tabular}[c]{@{}l@{}}Fee/Rate per \\ Transaction\end{tabular}}                                                & \textbf{\begin{tabular}[c]{@{}l@{}}Number of\\ Transactions\\ per Year\end{tabular}} & \textbf{\begin{tabular}[c]{@{}l@{}}Annual Gross\\ Revenues\end{tabular}} & \textbf{\begin{tabular}[c]{@{}l@{}}Annual Total\\ Fees\end{tabular}} \\ \hline
\textbf{Check Cashing}                                                 & \begin{tabular}[c]{@{}l@{}}2-3\% for payroll and\\ government checks\\ (Can exceed 15\% for\\ personal checks)\end{tabular} & 180 million                                                                          & \$60 billion                                                             & \$1.5 billion                                                        \\ \hline
\textbf{Payday Loans}                                                  & \begin{tabular}[c]{@{}l@{}}15-17\% per 2 weeks\\ 400\% APR\end{tabular}                                                     & 55-69 million                                                                        & \begin{tabular}[c]{@{}l@{}}\$10 - 13.8 \\ billion\end{tabular}           & \begin{tabular}[c]{@{}l@{}}\$1.6 - 2.2 \\ billion\end{tabular}       \\ \hline
\textbf{Pawnshops}                                                     & \begin{tabular}[c]{@{}l@{}}1.5 - 25\% monthly \\ 30-300\% APR\end{tabular}                                                  & 42 million                                                                           & \$3.3 billion                                                            & N/A                                                                  \\ \hline
\textbf{Rent-to-Own}                                                   & 2-3 times retail                                                                                                            & 3 million                                                                            & \$4.7 billion                                                            & \$2.35 billion                                                       \\ \hline
\textbf{\begin{tabular}[c]{@{}l@{}}Auto Title \\ Lenders\end{tabular}} & \begin{tabular}[c]{@{}l@{}}1.5 - 25\% monthly\\ 30-300\% APR\end{tabular}                                                   & N/A                                                                                  & N/A                                                                      & N/A                                                                  \\ \hline
\textbf{Total}                                                         & N/A                                                                                                                         & 280 million                                                                          & \$78 billion                                                             & \$5.45 billion                                                       \\ \hline
\end{tabular}
\end{table}

The growth of the AFS industry is one indicator and consequence of a large population being excluded from traditional financial services and forced to use exploitative financial services. The industry of alternative financial services has grown considerably over the past few decades. Michael Stegman of the University of North Carolina, Chapel Hill, reported that payday lending grew nationally from 300 stores seven years ago to more than 8,000 in 1999 (Carr and Schuetz, 2001). Further, in 1998, there were 7,500 rent-to-own stores that served three million customers, according to a Federal Trade Commission survey held in 2000 (Lacko et al. 2000). In fact, Norman D’Amours, the former chairman of the National Credit Union Administration, estimated that there are between 12,000 and 14,000 pawn shops across the country, outnumbering credit unions and banks (Carr and Schuetz, 2001). The unprecedented growth in the AFS industry has brought with it, unsurprisingly, unprecedented revenue as well. As shown in Table 8, check cashing service providers have made an annual total of \$60 billion in gross revenue and payday loan service providers have made around \$14 billion (Carr and Schuetz, 2001). It is important to note that this data does not represent recent trends - with the increasing influx of immigrants and the COVID-19 pandemic, the AFS industry is likely to see large growth in revenue. The growth in the revenue of AFS providers can be attributed to high fees and high APRs (Annual Percentage Rates), especially since deceptive interest rates and high-fee extensions trap unbanked Americans in a cycle of indebtedness. AFS providers exploit the low financial literacy and unfortunate circumstances of unbanked Americans to turn heavy profits. With banks unable to meet the needs of the unbanked and AFS providers exploiting their naivete in the financial world, the unbanked have no one to turn to. This is evident in a study by Sawyer which found that alternative service providers are concentrated in low-income areas with minority demographic populations - they prey on these unbanked and financially illiterate Americans and expand their services through revenue acquired through extremely high fees (Sawyer, 2004). In a policy brief exploring the lives of the unbanked in Buffalo, Jessica Gilbert conversed with multiple unbanked residents to gain a better understanding of how their life has been impacted by being unbanked. One resident said, “That’s the main thing that you do when you’re broke, you pawn your TV, right?” (Gilbert, 2018). Another stated, “You ever hear of the phrase, ‘the rich get richer and the poor stay poorer’?” (Gilbert, 2018). With no one but AFS providers to turn to, unbanked Americans lose hope in the financial system, ultimately leaving them with two options: refusal to use any financial services in the future or continuing to use predatory AFS. Unfortunately, both have detrimental consequences. 

The Buffalo policy brief detailed the consequences to both choices (Gilbert, 2018). Some unbanked Buffalo residents choose to store their money at home, “I was just thinking about it recently, you know, I work, I get checks and stuff so I want to be able to direct deposit, but I don’t want someone else spending my money for me”(Gilbert, 2018). Many unbanked Americans share similar concerns, but storing money at home is not a better option. One resident’s brother lost \$900 when his house caught on fire (Gilbert, 2018). Although rare, emergencies can cause the unbanked to lose all of their savings, ultimately putting them and their families in a vulnerable position. Rent-to-own stores, a type of AFS provider, are no better, as they leech money from the unbanked through expensive extensions. Rent-to-own stores lease furniture, appliances, electronics, and other items to customers through periodic payments (Gilbert, 2018). Customers can choose to stop making payments if they no longer need the item, or complete the payment schedule to become the owner of the item they are leasing.
There are three issues that the unbanked face here. The first is that if they complete all the payments required for ownership, they end up having to pay 2 or 3 times more than the interest included in their periodic payments (Gilbert, 2018). In fact, 60-70\% of customers purchase their leased items (Gilbert, 2018). With these “hidden” fees through interest, customers are effectively paying high prices for low-quality items. Secondly, while customers do not need to have access to credit to lease items, the rent-to-own stores do not report on-time payments to credit bureaus to help customers have a chance of having access to credit. Third, if customers miss a single payment, stores repossess their item, so the customer loses the item as well as all previous payments made towards the item (Gilbert, 2018). In short, no matter what path a customer takes, they will face a loss. What is most surprising is how such fees build up over time, ultimately forming a large divide between fees paid by the banked and the unbanked over a long period of time. 

In talking about the effects of using AFS, it’s also important to explore the costs associated with not being banked - the benefits of bank accounts that are unavailable to the unbanked. The benefits of bank accounts include savings for retirement, emergency funds, and cushion money, among other things. Additionally, the unbanked don’t have access to credit - they have to take out high-interest personal loans or pay out of pocket in these cases. Without savings accounts, unbanked Americans have little to contribute to their retirement each year, ultimately forcing them to work in their old age. Further, let's look at the cost comparison between traditional bank accounts and highly-popular prepaid cards. Traditional bank accounts cost \$7 a month, while community credit union fees and deposits amount to \$0. On the other hand, the minimum fees for a Walmart Money Card amount to \$12, and a GreenDot prepaid card amounts to \$20.80. In the long run, the benefits of bank accounts outweigh the benefits of AFS, while the costs of AFS are consistently higher than those of traditional bank accounts (Breitbach, 2003). 

\section{Solutions: Banking the Unbanked and Improving Financial Literacy}

This section will focus on banking the unbanked and improving financial literacy. The solutions that I believe will bank the unbanked are those solutions that have been implemented and shown tangible results through studies or experiments. The first solution we will investigate is the South African E Bank Model by Standard Bank, which proved successful for the unbanked in South Africa (Freund and Weil, 1999).  The most important step Standard Bank took was to conduct market research to accurately assess the needs of the unbanked in order to design a product to meet their needs. In designing a product for the unbanked, the most crucial feedback and ideas will come from the unbanked. Banks lose out on a large part of the market either because they ignore the needs of the unbanked or cannot accurately assess their needs. Market research discovered the particular needs of these low-income or unbanked customers: there was an urgent need for superior user-friendliness, faster speed of transaction, greater convenience, and a high degree of safety and security for the transactor. Standard Bank set up an independent electronic card-based banking operation to engage with this market. Customers’ biometric identification and facial image were stored in a central database to increase security and safety. Highly graphic illustrations and personal assistance increased user-friendliness. Lastly, hundreds of electronic branches were opened in low-income areas to increase convenience for potential customers. This solution was highly successful, with the bank converting over 2 million “book-based” low-income savers to a new card-based system. Over 75,000 new accounts are being opened every month (Freund and Weil, 1999).

It is important to look at exactly what made this initiative successful and how it could be adapted to the U.S. First, the card-based system was easy to understand. Second, there was high personal assistance for consumers, if needed. Third, mass marketing features like coupons and discounts served as incentives to attract more consumers (Freund and Weil, 1999).  The main issue is to carefully develop a plan to improve mass marketing and incentivize more unbanked Americans to use electronic banking and offer new, but limited services. Most unbanked Americans are already familiar with the technology used (McDonald’s and Taco Bell use similar kiosks for customers to order food) and high personal assistance is unnecessary because of America’s high literacy rates. With such advancements, I strongly believe in the electronic card-based system to bank the unbanked. However, it is important to gradually expand the market after perfecting the implementation of this situation, while continually conducting market research to orient solutions towards the needs of unbanked Americans. With many unbanked Americans likely to distrust initiatives by banks, it is crucial that the plan be implemented gradually to prevent mistakes after the initiative reaches the whole market, as this may lead to further divide and distrust between the unbanked and financial institutions.

Another solution to banking the unbanked that has become increasingly popular in countries like Kenya is using mobile banking as a channel for financial inclusion. Kenya is a leader in mobile payments implementation and adoption (Gross et al.  2012). As per the World Bank Findex Data, 60 percent of Kenyan adults over the age of 15 use mobile payments to send money, and 66 percent use mobile payments to receive money (Demirguc-Kunt and Klapper, 2012). Additionally, in the 144 countries that were surveyed, Kenya’s use of mobile financial services was 20 percentage points higher than in any other country (Gross et al. 2012). Further, mobile financial services have been proven to reach over 43\% of the unbanked population in India. Since these solutions work in the more rural environments of African countries and India, there shouldn’t be a reason why they shouldn’t work in the United States. The U.S has higher literacy rates, a better standard of living, better access to education, and better infrastructure.

In the U.S., in spite of unbanked residents having relatively high access to mobile banking, they choose not to engage in mobile banking. A Federal Reserve study reported that 89.7\% of fully banked, 91.4\% of underbanked, and 63.4\% of unbanked respondents had access to mobile phones (The Federal Reserve System, 2012). Mobile banking has been popular in multiple countries and thus, is a viable and easy-to-use method for using traditional banking services.  The U.S. government could therefore implement policies to incentivize the unbanked to use mobile banking. It’s hard to know which incentives would appeal the most to unbanked Americans, but like all innovative solutions, market research is definitely the first step to take in order to encourage financial inclusion through mobile banking. 

As shown in the last two solutions, meeting the needs of the unbanked should be the number one priority for bringing the unbanked into the financial mainstream. This innovative solution utilizes the emergence of card-based and prepaid systems, coupled with the explosive growth of cellular technology, to transform the way the unbanked access and use financial services (Prior and Santomá, 2010). Essentially, it implements “transformational business models” of mobile banking using the already existing prepaid platforms the unbanked are familiar with that have been implemented in the Philippines and in South Africa. DoCoMo has been highly successful in mobile banking in Japan by persuading large financial institutions to invest in the unbanked market through attractive financial terms (Prior and Santomá, 2010). This solution is likely to be successful in the United States because it uses unbanked Americans’ familiarity with prepaid cards. It would not require the unbanked to invest time or money to become familiar with a different system, enabling them to continue using the services they are more familiar with while providing them the security and safety a bank account provides. Instead of transitioning them directly from AFS to bank accounts - a huge change for many unbanked Americans - this solution most accurately meets the needs of the unbanked and transitions them into joining the financial mainstream rather than pushing them directly into the mainstream with no knowledge or experience to make sound financial decisions. Additionally, the unbanked can even use this solution as an opportunity to gain experience in using services that are more similar to traditional banking services, thus giving them the opportunity and access to traditional bank accounts and financial services if they are interested. 

Consumers can reload their cards easily online or by phone. Additionally, consumers have access to their monthly purchase statements and transaction history at any time, reducing the chance they are caught off-guard by “hidden fees” or similar costs. Finally, let’s look at features that are useful in the development of low-cost payment systems that respond to the reasons the unbanked have cited as to why they don’t use traditional banking services. Unbanked Americans will not face minimum balance requirements, and can access their financial transaction history and receive financial statements at any time (Prior and Santomá, 2010). Additionally, this payment system is designed for micropayments and microdeposits.

With the ability to reload and load prepaid cards at convenient locations near their residence, unbanked Americans benefit immensely from prepaid cards. Secondly, prepaid cards lack identification requirements and credit requirements that bar millions of individuals across the country from using bank accounts. Many immigrants stated issues with identification requirements and the time it took for their accounts to be authorized (Federal Reserve Bank, 2010). Thus, prepaid cards will help remove a significant barrier many unbanked Americans face. Lastly, prepaid cards are very hard to overdraft, resulting in minimal to no fees.

Another popular solution includes partnerships between banks and fringe lenders/AFS providers in addition to community credit unions providing products and services that best meet the needs of the unbanked in the United States. These partnerships are proving very beneficial to the unbanked in bad financial situations by offering them reasonable prices and savings vehicles in order to have money saved for their future (Carr and Schuetz, 2001). Bank One and the Chicago CRA coalition cited by Stegman, in chapter 8 of \textit{Banking the Unbanked}, are just some examples of these partnerships (Sherraden, 2005). Their pilot program, “Alternative Banking program”, aims to promote deposit services to unbanked consumers through offering safe, convenient, and inexpensive alternatives to check-cashing services while also conducting financial literacy workshops to demonstrate cost comparisons of financial services (Carr and Schuetz, 2001). These workshops dissuade unbanked Americans from using high-interest AFS. These partnerships provide many benefits for Chicago residents through lending, service, and investments in lower-income communities. Today, unbanked residents are either denied loans and other financial services because of the potential “high-risk” of investing in the unbanked or are forced to use AFS, where they are financially exploited. With products aimed at aiding the unbanked to help them reach a state of financial stability through joining the financial mainstream, these programs are not just providing lifelines to save unbanked Americans but rather providing them with the skills and knowledge to be financially independent in the future. 

Other innovative solutions include establishing community development credit unions (CDCUs) and community development financial institutions (CDFIs) in low-income communities that offer affordable alternatives while also providing the unbanked with valuable experiences in using financial services (Carr and Schuetz, 2001).  This has the effect of encouraging the unbanked to use traditional banking services and mainstream credit services as they provide similar services. Both credit unions offer low annual interest rates of 17\%-18\% with \$15 and \$30 processing fees and timely repayment requirements (Carr and Schuetz, 2001). These fees and interest rates serve two purposes: to fund the community credit unions to increase the number of unbanked reached and provide them with the experience needed to use traditional financial services in the future.

In general, solutions to financial exclusion rely on both profit-seeking institutions and nonprofits. Large financial institutions rely on these unbanked Americans learning how to use their services and pay their fees on time in order to make a profit. This means helping the unbanked is in their interest.  One example of this solution has been implemented by the Union Bank of California. They provided a lower-cost product through a division of their bank called Cash \& Save. It offers cash-checking services at a significantly lower percent fee - 1 to 1.5 percent fee - on payroll checks in the area (Carr and Schuetz, 2001). Another example is Direct Inc. – they offer a low-cost wire service that transfers money to foreign bank accounts at a very low cost (Carr and Schuetz, 2001). This enables immigrants to send money back home to their families, while saving on costs, and providing their families with easier access to the money. The critical issue the US needs to address is connecting unbanked Americans to these innovative services by building trust between unbanked consumers and big financial institutions. Thus, these innovative solutions can only be implemented with a three-pronged approach by the government, for-profit financial institutions, and non-profits (Carr and Schuetz, 2001).
First, the government has the regulatory authority to reduce the power and number of AFS providers in the US. Banks provide financial services and education to the unbanked. The non-profit agencies act as a bridge, connecting the unbanked with banks because these agencies have accumulated the trust of this population over many years. These solutions will only work if all three prongs work in conjunction - a banking initiative relies on the support of all three entities in order to be successful. 

This approach has been successfully implemented by the North Carolina State Employees’ Credit Union (SECU), as shown in a study by Stegman on payday lending (Stegman, 2007).  They modified an existing open-end line of consumer credit to create the Salary Advance Loan. Those whose paychecks were on direct deposit, had not caused the credit union losses in the past, and were not in bankruptcy were eligible for a SALO up to \$500. The monthly APR is around 12\% (Stegman, 2007). The results were very encouraging: over 2/3 of all SALO customers took advances nearly every month. Additionally, only 1.4\% of customers were 60 days delinquent while 0.65\% were 90 days delinquent. In comparison to 25\% of pawn shop customers defaulting and 20\% of car-title borrowers defaulting, delinquency rates for SALO customers are extremely low. Additionally, the charge-off percentages (around 0.2\%) serve to prove that the SALO customers do pay their loans off on time. The implication is that, if provided with fair loan terms, the risk of defaulting or delinquency on loans is minimal for unbanked Americans. 

Finally, it is important to address solutions to improving financial literacy in the United States. First, let’s look at previous literature on programs implemented to improve financial literacy because there are mixed results. One study by Mandell in 2008 used data from the Jump\$tart bi-annual financial literacy tests to support the conclusion that financial education is not very effective (Mandell, 2008). Another study on the effectiveness of financial education, however, makes the claim that the research that disproves the effectiveness of financial education is based on surveys with multiple structural errors including inefficient controls in course content, teacher preparation, knowledge in the curriculum, test measurement, and more (Walstad, 2010). This extensive study on financial literacy analyzed financial literacy test scores in middle and high-schoolers after being taught based on one curriculum: the FFL (Financial Fitness for Life) curriculum. Teachers participated in FFL workshops and thus, consistency in course content and curriculum was much higher. The test reliability was .86 (very high). High school teachers in three states administered this test to over 524 students (taught the FFL curriculum) and 335 similar students (control group). These changes (the study controlled for multiple factors that the Jump\$tart study did not) resulted in positive growth between pre-tests and post-test scores and a significant different between the control group and FFL group in terms of their average test scores. Additionally, there have been multiple studies conducted to measure the effectiveness of the FFL curriculum and they corroborate this study (Walstad, 2010). You may be wondering, “It is obvious that students taught in the FFL curriculum will do better on the FFL test. How do we know that they are more financially literate?”. The answer: the FFL curriculum was developed by the Council for Economic Education  - a nationally accredited organization that has created multiple personal finance standards for the United States in addition to helping advise the United States government on policy changes to better financial literacy education for high school and middle school students. 

\section{Conclusion}

This paper has reviewed at-risk demographics for being unbanked/financially illiterate and the causes, effects, and solutions for financial illiteracy/being unbanked. The main finding is that those with lower education, minority races, lower income, and immigrants are more likely to be financially illiterate. Another main finding is that high fees, banking needs being met, no incentive or ability to save, and distrust of banks are causes for the unbanked not using traditional banking services. Additionally, unbanked Americans face extremely high fees, financial exploitation by AFS providers, predatory interest rates, and financial vulnerability. Finally, solutions like electronic card-based systems, mobile banking as a channel for financial inclusion, products that use existing AFS platforms the unbanked are familiar with, partnerships between the three entities (banks, non-profits, and the government), and properly implemented FFL curriculum help increase financial inclusion and literacy. 

\section*{Acknowledgements\centering}
I would like to express my deepest gratitude to Laura Nicolae. She is a Harvard Economics Ph.D. candidate and her guidance helped me throughout all stages of my research paper. I could not have completed this journey without her expertise in Economics research. 
\vspace*{0.5cm}

\section*{Appendix}

\begin{figure}[H]
\setcounter{figure}{11}.
\centering
\includegraphics[scale=0.4]{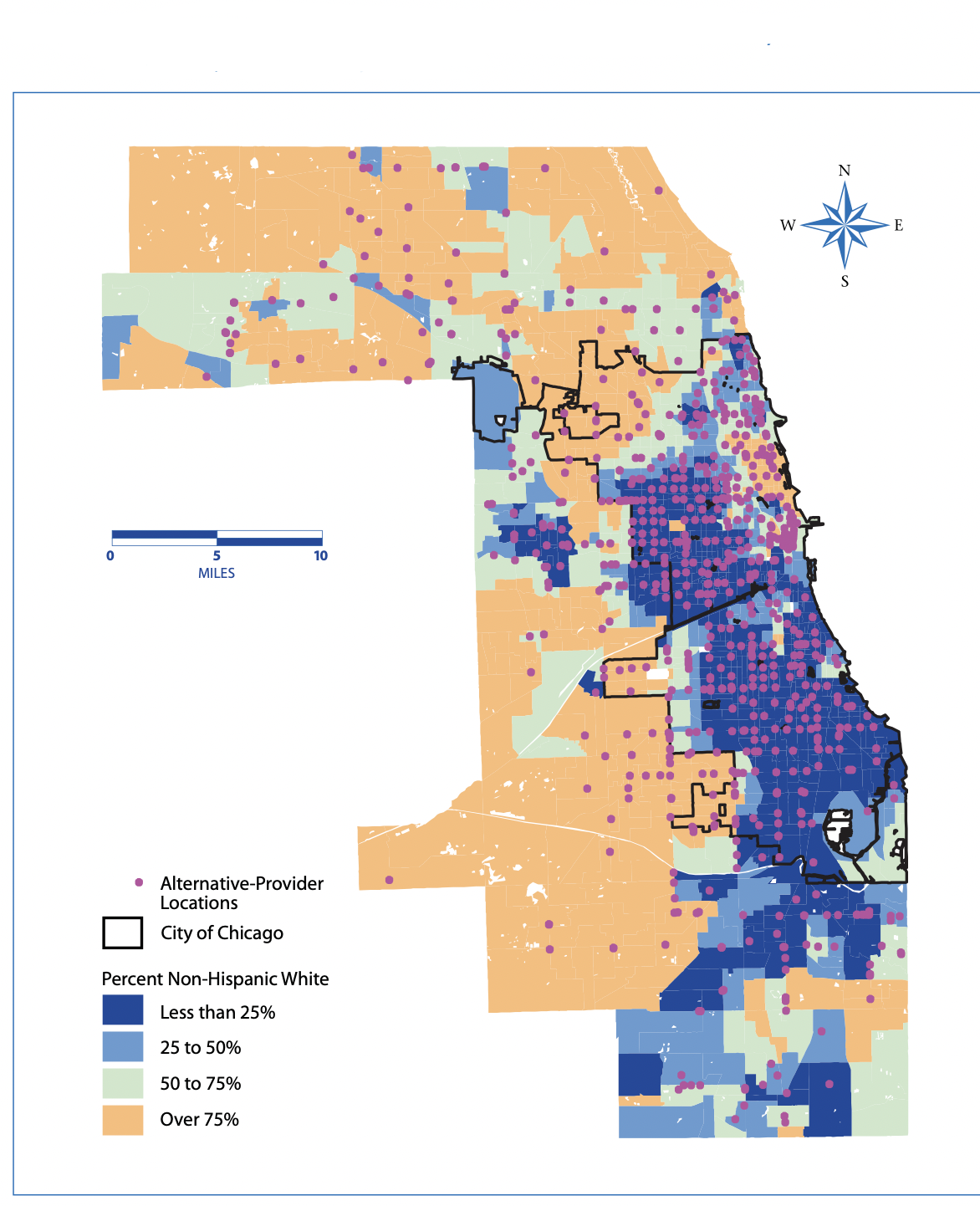}
\caption{Cook County, Illinois: Location of Alternative Financial Service Providers, With Percent Non-Hispanic White}
\label{fig:3}
\end{figure}

Figure 12 visually depicts the concentration of AFS providers in areas with low-income and minority populations.

\begin{figure}[H]
\centering
\caption{Alternative Financial Service Providers}
\includegraphics[scale=0.3]{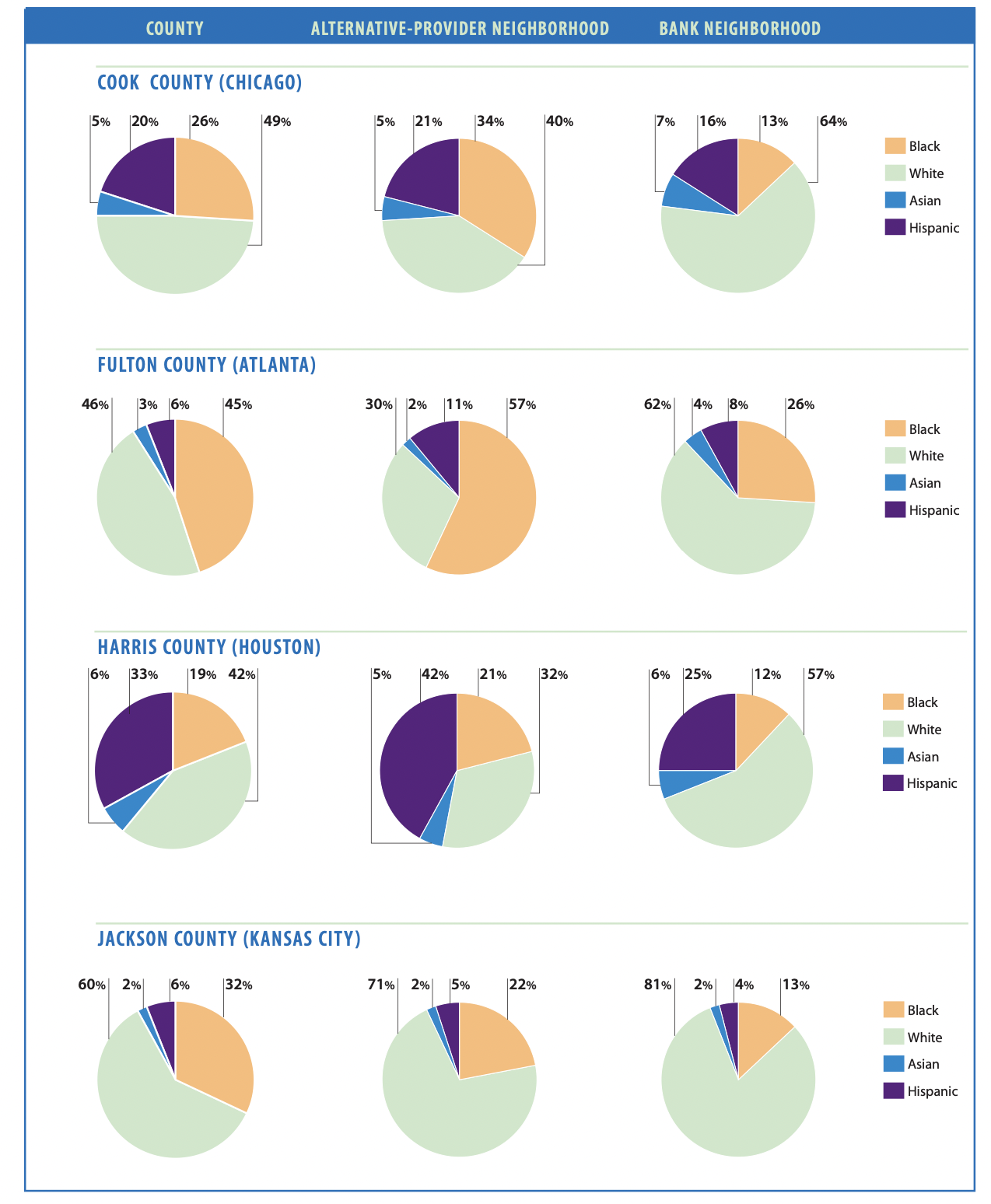}
\label{fig:3}
\end{figure}

\begin{figure}[H]
\centering
\caption{Alternative Financial Service Providers}
\includegraphics[scale=0.3]{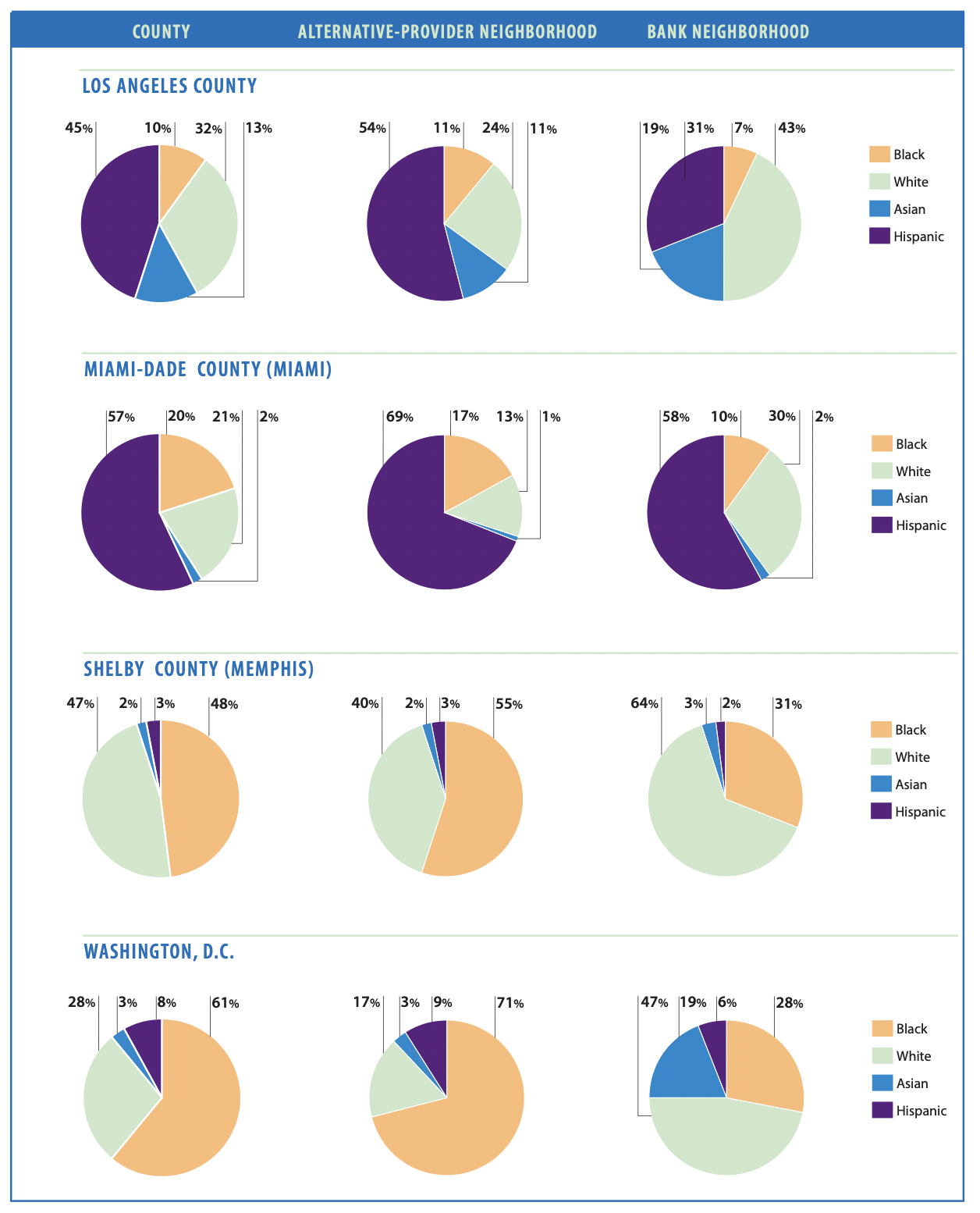}
\label{fig:3}
\end{figure}

The pie charts for each site show the racial/ethnic makeup of the county as a whole, the typical neighborhood of an alternative provider, and the typical neighborhood of a conventional bank.

\begin{figure}[H]
\setcounter{table}{14}.
\centering
\captionsetup[table]{skip=3pt}
\captionof{table}[foo]{Overview of North Carolina State Employees Credit Union (SECU) Salary Advance Loan (SALO) product}
\includegraphics[scale=0.7]{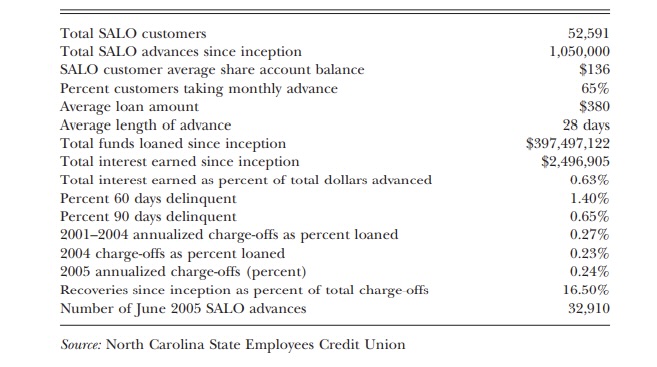}
\label{fig:3}
\end{figure}

The figure presents data on the success of the SALO product implemented by North Carolina State Employees Credit Union (SECU).

\end{document}